\documentclass[preprint]{revtex4}
\input epsf
\usepackage{amsmath,amssymb}
\usepackage{graphicx}

\draft

\begin{document}
\titlepage
\title{Palatini formulation of  the $R^{-1}$modified gravity with an additionally squared
 scalar curvature term}
\author{Xinhe Meng$^{1,2,3}$ \footnote{xhmeng@phys.nankai.edu.cn}
 \ \ Peng Wang$^2$ \footnote{pewang@eyou.com}
} \affiliation{1. CCAST(World Laboratory), P.O.Box 8730, Beijing 100080, China
\\2.  Department of Physics, Nankai University,
Tianjin 300071, P.R.China \\3. Institute of Theoretical Physics,
CAS, Beijing 100080, P.R.China
}

\begin{abstract}
In this paper by deriving the Modified Friedmann equation in the
Palatini formulation of $R^2$ gravity,  first we  discuss
the problem of whether in Palatini formulation an additional $R^2$ term 
in Einstein's General Relativity action can
drive an inflation. We show that the Palatini formulation of $R^2$
gravity cannot lead to the gravity-driven inflation as in the metric formalism. If considering no
zero radiation and matter energy densities, we obtain that only
under rather restrictive assumption about the radiation and matter
energy densities there will be a mild power-law inflation $a(t)\sim
t^2$, which is obviously different from the original vacuum energy-like driven inflation. Then we demonstrate 
that in the Palatini formulation of a more generally modified
gravity, i.e., the $1/R+R^2$ model that intends to explain both
the current cosmic acceleration and early time inflation, accelerating 
cosmic expansion achieved at late Universe evolution times under the model
parameters satisfying $\alpha\ll\beta$.
\end{abstract}

\maketitle

\textbf{1. Introduction}

Athough the fact that the expansion of our universe is currently in an accelerating
phase now seems well-established \cite{Perlmutter}, so far  the
mechanism responsible for this is yet not very clear. Many authors
introduce a mysterious cosmic fluid called dark energy  in General Relativity 
and Roberson-Walker metric framework to explain
this. See Ref.\cite{Carroll2} for a review and Ref.\cite{Dark} for
some recent models.

On the other hand, some authors suggested that maybe there does
not exist such a mysterious dark energy, but the observed cosmic
acceleration is a signal of our first real lack of understanding
of gravitational physics \cite{Lue, Carroll}. An example is the
braneworld theory of Dvali et al. \cite{Dvali}. Recently, some
authors proposed to add a $1/R$ term in the Einstein-Hilbert
action to modify the General Relativity (GR) \cite{Carroll,
Capozziello}. It is interesting that such a term may be predicted
by string/M-theory \cite{Odintsov3}. In the metric formulation,
this additional term will give fourth order field equations. It
was shown in their works that this additional term can give
accelerating expansion solutions for  the field equations without
dark energy. In this framework, Dick \cite{Dick} considered the
problem of weak field approximation, and Soussa and Woodard
\cite{Woodard} considered the gravitational response to a diffuse
source.

Based on this modified action, Vollick \cite{Vollick} has used
Palatini variational principle to derive the field equations. In
the Palatini formulation, instead of varying the action only with
respect to the metric, one views the metric and connection as
independent field variables and vary the action with respect to
them independently. This would give second order field equations.
For the original Einstein-Hilbert action, this approach gives the
same field equations as the metric variation. For a more general
action, those two formulations are inequivalent, they will lead to
different field equations and thus describe different physics
\cite{Volovich}. Flanagan \cite{Flanagan} derived the equivalent
scalar-tensor description of the Palatini formulation. In
Ref.\cite{Dolgov}, Dolgov and Kawasaki argued that the fourth
order field equations in metric formulation suffer serious
instability problem. If this is indeed the case, the Palatini
formulation appears even more appealing, because the second order
field equations in Palatini formulation are free of this sort of
instability \cite{Wang}. Furthermore, Chiba \cite{Chiba} argued
that the theory derived using metric variation is in conflict to
the solar system experiments. However, the most convincing
motivation to take the Palatini formalism seriously is that the
Modified Friedamnn (MF) equation following from it fit the SN Ia
data at an acceptable level \cite{Wang}.

On the other end of cosmic time, the very early stage, it is now
generally believed that the universe also undergos an acceleration
phase called inflation. The mechanism driven inflation is also
unclear now. The most popular explanation is that inflation is
driven by some inflaton field \cite{Liddle}. Also, some authors
suggest that modified gravity could be responsible for inflation
\cite{Starobinskii, Odintsov}. Revealing the mechanisms for
current acceleration and early inflation are two of the most
important objects of modern cosmology.

As originally proposed by Carroll et al. \cite{Carroll} and later
implemented by Nojiri and Odintsov \cite{Odintsov}, adding
correction term $R^m$ with $m>0$ in addition to the $1/R$ term may
explain both the early time inflation and current acceleration
without by introducing inflaton and dark energy. Furthermore, Nojiri and Odintsov
\cite{Odintsov} showed that adding a $R^m$ term can avoid the
above mentioned instability when considering the theory in metric
formulation. In this paper, we will show that in the Palatini
formulation, the $R^2$ term contribution is not same as to the
conclusion when considering the theory in metric formulation
\cite{Starobinskii}. And the $1/R+R^2$ model will have some
theoretical inconsistencies as well as conflict with particle
experiments that might invalid this model.

Besides, there are many activities in the study of quantum versions of $R^2$
gravity which seems to be a multiplatively renormalizable theory
(for a review, see Ref.\cite{Buchbinder}). However, such theory
has had a serious problem: possible non-unitarity due to the presence of
higher derivative terms. It is very promising that in Palatini
formalism higher derivative terms do not play such a role as in
metric formalism such that the unitarity problem of $R^2$
gravity may be resolved in Palatini formalism.
Aslo, it is interesting to explore the $R^2$ correction to the
chaotic inflation scenario \cite{R2} in Palatini formulation. When
written in Einstein frame, in metric formulation, this will
correspond to two scalar field inflation; in the Palatini
formulation, the model will correspond to a type of k-inflation
\cite{k-inflation}. More detailed investigations of this idea can
be found in our recently published work\cite{mg}.

This paper is arranged as follows: in Sec.2 we review the
framework of deriving field equations and Modified Friedmann (MF)
equations in Palatini formulation; in Sec.3 we discuss $R^2$
gravity in Palatini formulation and show the cosmology implyings;
in Sec.4 we discuss the combined effects of both a $1/R$ term and
a $R^2$ term; Sec.5 is devoted to conclusions and discussions.

\textbf{2. Deriving the Modified Friedmann equation in Palatini
formulation}

Firstly, we briefly review deriving field equations from a
generalized Einstein-Hilbert action by using Palatini variational
principle. See refs. \cite{Vollick, Wang, Volovich} for details.

The field equations follow from the variation in Palatini approach
of the generalized Einstein-Hilbert action
\begin{equation}
S=-\frac{1}{2\kappa^2}\int{d^4x\sqrt{-g}L(R)}+S_M\label{action}
\end{equation}
where $\kappa^2 =8\pi G$, $L$ is a function of the scalar
curvature $R$ and $S_M$ is the matter action.

Varying with respect to $g_{\mu\nu}$ gives
\begin{equation}
L'(R)R_{\mu\nu}-\frac{1}{2}L(R)g_{\mu\nu}=\kappa^2
T_{\mu\nu}\label{2.2}
\end{equation}
where a prime denotes differentiation with respect to $R$ and
$T_{\mu\nu}$ is the energy-momentum tensor given by
\begin{equation}
T_{\mu\nu}=-\frac{2}{\sqrt{-g}}\frac{\delta S_M}{\delta
g^{\mu\nu}}\label{2.3}
\end{equation}
We assume the universe contains dust and radiation, thus
$T^{\mu}_{\nu}=\{-\rho_m-\rho_r,p_r,p_r,p_r\}$ where $\rho_m$ and
$\rho_r$ are the energy densities for dust and radiation
respectively, $p_r$ is the pressure of the radiation. Note that
$T=g^{\mu\nu}T_{\mu\nu}=-\rho_m$ because of the relation
$p_r=\rho_r/3$.

In the Palatini formulation, the connection is not associated with
$g_{\mu\nu}$, but with $h_{\mu\nu}\equiv L'(R)g_{\mu\nu}$, which
is known from varying the action with respect to $\Gamma
^{\lambda}_{\mu\nu}$. Thus the Christoffel symbol with respect to
$h_{\mu\nu}$ is given by
\begin{equation}
\Gamma
^{\lambda}_{\mu\nu}=\{^{\lambda}_{\mu\nu}\}_g+\frac{1}{2L'}[2\delta
^{\lambda}_{(\mu}\partial
_{\nu)}L'-g_{\mu\nu}g^{\lambda\sigma}\partial
_{\sigma}L']\label{Christoffel}
\end{equation}
where the subscript $g$ signifies that this is the Christoffel
symbol with respect to the metric $g_{\mu\nu}$.

The Ricci curvature tensor is given by
\begin{eqnarray}
R_{\mu\nu}=R_{\mu\nu}(g)+\frac{3}{2}(L')^{-2}\nabla _{\mu}L'\nabla
_{\nu}L' -(L')^{-1}\nabla _{\mu}\nabla
_{\nu}L'-\frac{1}{2}(L')^{-1}g_{\mu\nu}\nabla _{\sigma}\nabla
^{\sigma}L'\label{Ricci}
\end{eqnarray}
and
\begin{equation}
R=R(g)-3(L')^{-1}\nabla _{\mu}\nabla ^{\mu}
L'+\frac{3}{2}(L')^{-2}\nabla_{\mu}L'\nabla^{\mu}L'\label{scalar}
\end{equation}
where $R_{\mu\nu}(g)$ is the Ricci tensor with respect to
$g_{\mu\nu}$ and $R=g^{\mu\nu}R_{\mu\nu}$. Note by contracting
(\ref{2.2}), we get:
\begin{equation}
L'(R)R-2L(R)=\kappa^2 T\label{R(T)}
\end{equation}
Assume we can solve $R$ as a function of $T$ from (\ref{R(T)}).
Thus (\ref{Ricci}), (\ref{scalar}) do define the Ricci tensor with
respect to $h_{\mu\nu}$.

Then we review the general framework of deriving modified
Friedmann equation in Palatini formalism \cite{Wang}. Let us work
with the Robertson-Walker metric describing the cosmological
evolution,
\begin{equation}
ds^2=-dt^2+a(t)^2(dx^2+dy^2+dz^2)\label{metric}
\end{equation}
Note that we only consider a flat metric, which is favored by
present observations \cite{Perlmutter}.

From (\ref{metric}), (\ref{Ricci}), we can get the non-vanishing
components of the Ricci tensor:
\begin{equation}
R_{00}=-3\frac{\ddot{a}}{a}+\frac{3}{2}(L')^{-2}(\partial_0{L'})^2-\frac{3}{2}(L')^{-1}\nabla_0\nabla_0L'\label{R00}
\end{equation}
\begin{eqnarray}
R_{ij}=[a\ddot{a}+2\dot{a}^2+(L')^{-1}\{^0_{ij}\}_g\partial_0L'
+\frac{a^2}{2}(L')^{-1}\nabla_0\nabla_0L']\delta_{ij}\label{ij}
\end{eqnarray}

Substituting equations (\ref{R00}) and (\ref{ij}) into the field
equations (\ref{2.2}), we can get
\begin{equation}
6H^2+3H(L')^{-1}\partial_0L'+\frac{3}{2}(L')^{-2}(\partial_0L')^2=\frac{\kappa^2
(\rho+3p)+L}{L'}\label{aa}
\end{equation}
where $H\equiv \dot{a}/a$ is the Hubble parameter, $\rho$ and $p$
are the total energy density and total pressure respectively.
Assume that we can solve $R$ in term of $T$ from Eq.(\ref{R(T)}),
substitute it into the expressions for $L'$ and $\partial_0L'$, we
can get the MF equation.

In this paper we will consider the Palatini formulation of the
following model suggested by Carroll et al. \cite{Carroll} and
implemented in the metric formulation by Nojiri and Odintsov
\cite{Odintsov}:
\begin{equation}
L=R-\frac{\alpha^2}{3R}+\frac{R^2}{3\beta}\label{Rfull}
\end{equation}
where $\alpha$ and $\beta$ are parameters both with dimensions
$(eV)^2$.

Since in early universe, the $R^2$ term is dominated. In order to
find what this term functions, we first consider the Palatini
formulation of the modified action with only a $R^2$ term, that is:
\begin{equation}
L=R+\frac{R^2}{3\beta}\label{R2}
\end{equation}
This action has been studied by Starobinsky in metric formulation
\cite{Starobinskii} and it has been shown that a gravity-driven
inflation can be achieved.

\textbf{3. Palatini formulation of $R^2$ gravity}

The field equations follow by substituting Eq.(\ref{R2}) into
Eq.(\ref{2.2})
\begin{equation}
(1+\frac{2R}{3\beta})R_{\mu\nu}-\frac{1}{2}g_{\mu\nu}(R+\frac{R^2}{3\beta})=\kappa^2
T_{\mu\nu}\label{R2field}
\end{equation}

Contracting indices gives
\begin{equation}
R=-\kappa^2 T=\kappa^2\rho_m\label{RR}
\end{equation}
The second equality follows because the radiation has vanishing
trace of momentum-energy tensor. This equation is quite
remarkable, since it is formally the same as the one given by GR,
with only one difference: $R_{\mu\nu}$ is associated with the
conformal transformed matric $h_{\mu\nu}=L'(R)g_{\mu\nu}$ and
$R=g^{\mu\nu}R_{\mu\nu}$.

From the conservation equation $\dot{\rho_m}+3H\rho_m=0$ and
Eq.(\ref{RR}), we can find that
\begin{equation}
\partial_0L'=-2\frac{\kappa^2\rho_m}{\beta}H\label{}
\end{equation}

Substituting this into Eq.(\ref{aa}) we can get the Modified
Friedmann equation for the $R^2$ gravity:
\begin{equation}
H^2=\frac{2\kappa^2(\rho_m+\rho_r)+\frac{(\kappa^2\rho_m)^2}{3\beta}}{(1+\frac{2\kappa^2\rho_m}{3\beta})[6+3F_0(
\frac{\kappa^2\rho_m}{\beta})(1+\frac{1}{2}F_0(\frac{\kappa^2\rho_m}{\beta}))]}\label{R2MF}
\end{equation}
where the function $F_0$ is given by
\begin{equation}
F_0(x)=-\frac{2x}{1+\frac{2}{3}x}\label{F0}
\end{equation}
It is interesting to see from Eq.(\ref{R2MF}) that all the effects
of the $R^2$ term are determined by $\rho_m$. If $\rho_m=0$,
Eq.(\ref{R2MF}) simply reduces to the standard Friedmann equation.

Now let's come to the discussion of inflation. To begin with, note
that in the metric formulation of the $R^2$ gravity, inflation is
driven by the vacuum gravitational field, i.e. we assume that the
radiation and matter energy densities is zero during inflation,
thus called "gravity-driven" inflation. However, in the Palatini
formulation, when the radiation and matter energy densities is
zero, it can be seen directly from Eq.(\ref{R2MF}) that the
expansion rate will be zero and thus no inflation will happen.
Thus, in the Palatini formulation of $R^2$ gravity, we cannot have
a gravity-driven inflation. So the only hope that the $R^2$ term
can drive an inflation without an inflaton field is that the
relationship between the expansion rate and the energy density of
radiation and matter will be changed which can lead to inflation
(thus what we are talking now is similar to the "Cardassian"
scenario of Freese and Lewis \cite{freese}: the current
accelerated expansion of the universe is driven by the changed
relationship between the expansion rate and matter energy
density). We will see that naturally there will be no inflation
and a power-law inflation can happen only under specific
assumption on $\rho_m$ and $\rho_r$.

First, in typical model of $R^2$ inflation, $\beta$ is often taken
to be the order of the Planck scale \cite{Starobinskii}. This is
also the most natural value of $\beta$ from an effective field
point of view. Thus we naturally have $\kappa^2\rho_m/\beta \ll
1$. Under this condition, it can be seen that from Eq.(\ref{F0}),
we have $F_0\sim 0$, and the MF equation (\ref{R2MF}) reduces to
the standard Friedmann equation:
\begin{equation}
H^2=\frac{\kappa^2}{3}(\rho_m+\rho_r)\ .\label{st}
\end{equation}
Thus it is obvious that in this case there will be no inflation.
Also note that from the BBN constraints on the Friedmann equation
\cite{Carroll4}, $\beta$ should be sufficiently large so that the
condition $\kappa^2\rho_m/\beta \ll 1$ is satisfied at least in
the era of BBN. Thus we conclude that in the most natural case,
Palatini formulation of $R^2$ gravity cannot lead to inflation.

Second, let's assume that in the very early universe, we have
$\kappa^2\rho_m/\beta \gg 1$ that is the interesting possibility deviating GR. In this case, from Eq.(\ref{F0}),
the MF equation (\ref{R2MF}) will reduce to
\begin{equation}
H^2=\frac{\kappa^2\rho_m}{21}+\frac{2\beta\rho_r}{7\rho_m}+\frac{2\beta}{7}\
.\label{33}
\end{equation}
Then we can see that if the $\beta$ term could dominate over the
other two terms, it would drive an exponential expansion by the
effective cosmological constant $\beta$. But note that this
equation is derived under the assumption that $\beta\ll
\kappa^2\rho_m$. Thus inflation cannot be driven by the $\beta$
term. On the other hand, if we assume further that
$\rho_r\gg\kappa^2\rho_m^2/\beta$, i.e., the second term dominates
in the MF equation (\ref{33}) over some time interval if $\rho_{r0}^3>>\kappa^2\rho_{m0}^4/\beta$, 
and then if the matter and radiation evolve
independently so that from the relation
$\rho_r\propto a^{-4}$ and $\rho_m\propto a^{-3}$, the MF equation
(\ref{33}) can be solved to give $a(t)\propto t^2$ with neglecting
numerical factors. Thus, only in
this case, we can get a mild power-law inflation that quite differs from the original exponent inflation
 with enough e-folding for solving the Hot big bang cosmology
puzzles: lack of defects, flatness, horizon and homogeneous problems.  This kind of mild inflation 
will occur at a time smaller than the
timescale associated with $\beta^{-1/2}$ ( in this MF equation for a possible inflation there is a particular 
time scale associated to the parameter $\beta$  
being $\beta^{-1/2}$ by dimension anlysis), which may be unrealistic if the time
scale is of order of the Planck time. 
However, current
constraint with cosmic background radiation anisotropies 
(power spectum) analysis on the rate of power-law inflation reads $p>21$ where
$a\propto t^p$ (see, e.g., Ref.\cite{lim}). So this case is not a
viable model of inflation.

Besides, we can see it in another aspect with the e-folding number N large enough 
requried for solving the original cosmology problems.
\begin{equation}
N=ln[\frac{a_f}{a_i}]=2ln(t_f/t_i)\label{efd}
\end{equation}
If we ask the e-folding number $ N>60$ then we have $t_f>e^{30}t_i$, i.e, this kind of power-law inflation lasts 
not less than $10^{13}$ times the initial time which is against the primordial inflation required.
As additional comments we should mention that there are some low energy-scale
inflation taking place possibilities,
especially if one considers alternative ways to generate density
perturbations,
such as the curvaton mechanism or a modulated inflaton decay constant, but these scenaries 
are not relevant to what we focus on above.  

At late cosmological times when $\kappa^2\rho_m/\beta \ll 1$,
$F_0\sim 0$, the MF equation (\ref{R2MF}) reduces to the standard
Friedmann equation, which implies that early universe dynamics is dominated 
by larger curvature term and enlightens us to describe late times cosmologies 
by including possibly the sort of $R^{n}$  term with n as negative integer when the small curvature 
term dominates. Moreover as a reduction, this type of
inflation
also occurs in a more complicated model with an additional $R^{-1}$ term,
when taking this term's coupling constant going to zero directly, as we will discuss in next section below.

In summary, in the Palatini formulation, the modified gravity
theory with a $R^2$ correction term would not lead to an early 
time gravity-driven inflation, in opposite to the famous conclusion when
considering the theory in the metric formulation. The difference
of those two formulations is now quite obvious. Now, we still can
not tell which one is physical. But this makes those results more
interesting. It is conceivable that quantum effects of the $R^2$
theory in Palatini formulation would also be different from the
metric formulation (see Ref. \cite{Buchbinder} for a review). Such
higher derivative terms similar to $R^2$ term may be induced by
the quantum effects, e.g., trace anomaly \cite{Buchbinder,
Odintsov2}. It has been recently shown \cite{Odintsov2} that
phantom cosmology implemented by trace anomaly induced terms also
admits both early time inflation and late time cosmic
acceleration. It follows from our consideration that $R^2$ term in
Palatini formulation does not support inflation, then we expect that
also in phantom cosmology with quantum effects considered in Palatini
formulation, the inflation does not occur either.

\textbf{4. Palatini formulation of a $1/R+R^2$ gravity}

In this part we will consider mainly the cosmological consequences of
a $R^{-1} + R^{2}$ gravity theory, when this is analyzed in the
Palatini formulation.
The $R^{-1}$ phenomenological theory have gained some interest since
it seems to
be able to account for cosmological observations in Supernovae Ia, as an
alternative model to dark energy; this topic nowadays is a hot
subject under discussions \cite{cs}. Now
the additional $R^{-1}$ term is coupled to a $R^{2}$ term to investigate possibly
 more interesting cosmological features, such as
if the above discussed 
 kind of mild power-law with power 2 (early
Universe) inflation
could also happen. Qualitatively reasoning the large curvature terms' dominate cosmic 
global evolution at early times that the inverse curvature terms effect can be neglected 
for the early universe evolution stage, which is also reflected
in Inflation theory. In this case the above section discussions apply here.

 Now let's turn to discussions of the $1/R+R^2$ gravity more mathematically, especially 
 the relative strength for the two additional curvature terms.

The field equations follow by substituting Eq.(\ref{Rfull}) into
(\ref{2.2})
\begin{equation}
(1+\frac{\alpha^2}{3R^2}+\frac{2R}{3\beta})R_{\mu\nu}-\frac{1}{2}g_{\mu\nu}(R-\frac{\alpha^2}{3R}+\frac{R^2}{3\beta})=\kappa^2
T_{\mu\nu}\label{Rfullfield}
\end{equation}

Contracting indices gives explicit expression for scalar curvature
\begin{equation}
R=\frac{1}{2}\alpha [-\frac{\kappa^2
T}{\alpha}+2\sqrt{1+\frac{1}{4}(\frac{\kappa^2
T}{\alpha})^2}]=\frac{1}{2}\alpha [\frac{\kappa^2
\rho_m}{\alpha}+2\sqrt{1+\frac{1}{4}(\frac{\kappa^2
\rho_m}{\alpha})^2}]\label{RRfull}
\end{equation}
where we take the plus sign in the two root solution as the matter density is positive,
 and as in Sec.3, we assume the universe contains dust and
radiation. It is interesting to note that Eq.(\ref{RRfull}) is the
same as the one in $1/R$ gravity \cite{Wang}.

From the conservation equation $\dot{\rho_m}+3H\rho_m=0$ and
Eq.(\ref{RRfull}), we can find that
\begin{equation}
\partial_0L'=\frac{(\frac{\alpha}{R})^2-\frac{R}{\beta}}{\sqrt{1+\frac{1}{4}(\frac{\kappa^2\rho_m}{\alpha})^2}}
\frac{\kappa^2\rho_m}{\alpha}H\label{}
\end{equation}

Substituting this into Eq.(\ref{aa}) we can get the MF equation:
\begin{equation}
H^2=\frac{\kappa^2\rho_m+2\kappa^2\rho_r+\alpha[G(\frac{\kappa^2\rho_m}{\alpha})-\frac{1}{3G(\frac{\kappa^2\rho_m}{\alpha})}
+\frac{\alpha}{3\beta}
G(\frac{\kappa^2\rho_m}{\alpha})^2]}{[1+\frac{1}{3G(\frac{\kappa^2\rho_m}{\alpha})^2}+
\frac{2\alpha}{3\beta}
G(\frac{\kappa^2\rho_m}{\alpha})][6+3F(\frac{\kappa^2\rho_m}{\alpha})(1+\frac{1}{2}F(\frac{\kappa^2\rho_m}{\alpha}))]}
\label{fullMF}
\end{equation}
where the two functions $G$ and $F$ are given by
\begin{equation}
G(x)=\frac{1}{2}[x+2\sqrt{1+\frac{1}{4}x^2}]\label{G}
\end{equation}
\begin{equation}
F(x)=\frac{(1-\frac{\alpha}{\beta}G(x)^3)x}{(G(x)^2+\frac{2\alpha}{3\beta}G(x)^3+\frac{1}{3})\sqrt{1+\frac{1}{4}x^2}}
\label{F}
\end{equation}

In order to be consistent with observations, we should have
$\alpha\ll\beta$. We can see this in two different ways.

Firstly, when $\kappa^2\rho_m\gg\alpha$, from Eq.(\ref{G}),
$G\sim\kappa^2\rho_m/\alpha$. From the BBN constraints, we know
the MF equation should reduce to the standard one in the BBN era
\cite{Carroll4}. This can be achieved only when $F\sim0$ \cite{Wang} and from
Eq.(\ref{F}), this can be achieved only when $\alpha\ll\beta$ and
$1\ll\kappa^2\rho_m/\alpha\ll(\beta/\alpha)^{1/3}$.

Secondly, when $\kappa^2\rho_m\ll \alpha$ (this is just the case we are interested in 
for considering the deviation from GR), we can expand the
r.h.s. of Eq.(\ref{fullMF}) to the first order in
$\kappa^2\rho_m/\alpha$:
\begin{equation}
H^2=\frac{\frac{11+\alpha/\beta}{8+4\alpha/\beta}\kappa^2\rho_m+\frac{3}{2+\alpha/\beta}\kappa^2\rho_r+\frac{1}{2}\alpha}
{6+\frac{9}{4+2\alpha/\beta}(1-\alpha/\beta)\frac{\kappa^2\rho_m}{\alpha}}\label{1stfull}
\end{equation}

When $\alpha\ll\beta$, this will reduces exactly to the first
order MF equation in the 1/R theory \cite{Wang}. Since we have
shown there that the MF equation in 1/R theory can fit the SN Ia
data at an acceptable level, the above MF equation can not deviate
from it too large, thus the below condition  should be
satisfied consistently,
\begin{equation}
\alpha\ll\beta
\end{equation}
The coupling constant $\alpha$ is very small and the inverse curvature term functions only at larger 
cosmic scale as expected, which also can be seen from the recent analysis by Carroll et al \cite{Carroll, cs}.

Following Ref.\cite{Flanagan}, we have an equivalent scalar-tensor
description of the Palatini formulation of modified gravity. When
considering the $1/R+R^2$ gravity, the potential is given by
\cite{Chiba, Flanagan}
\begin{equation}
V(\Phi)=\frac{\frac{2\alpha^2}{3\phi}+\frac{\phi^2}{3\beta}}{2\kappa^2}\exp(-2\sqrt{
\frac{2\kappa^2}{3}}\Phi)\label{po2}
\end{equation}
where $\phi$ is determined from $\Phi$ by
\begin{equation}
\frac{2}{3}(\frac{\phi}{\beta})^3-(\exp(\sqrt{\frac{2\kappa^2}{3}}\Phi)-1)(\frac{\phi}{\beta})^2+
\frac{1}{3}(\frac{\alpha}{\beta})^2=0\label{phi}
\end{equation}

Now we can see a problem of $1/R+R^2$ gravity. The determinant of
Eq.(\ref{phi}) is
$\Delta=\frac{27}{4}(\frac{\alpha}{\beta})^2[(\exp(\sqrt{\frac{2\kappa^2}{3}}\Phi)-1)^3-(\frac{\alpha}{\beta})^2]$.
When it is positive, i.e. $\sqrt{\kappa^2}\Phi>
\sqrt{3/2}\ln(1+(\alpha/\beta)^{2/3})$, Eq.(\ref{phi}) has three
distinct real solutions. Thus the correspondence to scalar-tensor
theory is not one-to-one now. According to Ref.\cite{Magnano},
this is a strong indication that the $1/R+R^2$ theory is not a
consistent theory. Furthermore, as also pointed out to us by
\'{E}anna Flanagan \cite{foot}, this implies that when $R$ exceeds
the critical value $R_0=(\alpha^2/\beta)^{1/3}$ which satisfies
$L''(R_0)=0$, the $1/R+R^2$ theory has not a well-behaved
initial-value formulation.

In Ref.\cite{Flanagan}, Flanagan also showed that, if we assume
the $1/R$ model is also applicable at small scales, there will be
severe conflict with electron-electron scattering experiments.
This conflict is due to the smallness of the energy scale of the
potential near the extremal point, which is $\alpha/\kappa^2\sim
10^{-12} (ev)^4$ in the $1/R$ case. In the $1/R+R^2$ case, we can
find that the extremal value $V'(\Phi_0)=0$ is given by
$\sqrt{\kappa^2}\Phi_0=\sqrt{3/2}\ln(4/3+2\alpha/3\beta)$. This
corresponds to $\phi=\alpha$. Substitute this into Eq.(\ref{po2})
and use the fact $\alpha\ll\beta$ found above we can find that
near the extremal point $\Phi_0$ the potential is of the order
$\alpha/\kappa^2$, i.e. the same as the $1/R$ case. Thus the
conflict will still appear.

However, the above conflict is due to the fact that we assume the
$1/R$ corrected action is applicable in very small scales in
addition to the astrophysical scales where it is originally
suggested to be effective to explain the cosmic acceleration.
Thus, this $1/R$ gravity theory can not be a fundamental theory
and if we can find some way to guarantee that it is only effective
in large scale, we can still use it to discuss cosmological
issues. A concrete way to achieve this is still under
investigation.

\textbf{5. Conclusions and Discussions}

In this paper we have shown that in the Palatini formulation, a
$R^2$ term can not lead to an early time inflation, in opposite to
the conclusion when considering the theory in metric variation.
Furthermore, in the more general $1/R+R^2$ model that intends to explain both
the current cosmic acceleration and the early time inflation, we have demonstrated that
accelerating cosmology at late times can be obtained without dark energy introduced under the 
model coupling constants consistently satisfying the condition that $\alpha\ll\beta$.

Intuitively speacking, the cosmic global evolution at early times is dominated by large curvature
term like $R^{n}$  $ (n>0)$  while at later times by kind of $R^{n}$  $(n<0)$ term when the small curvature 
term dominates  the global evolution.
The current "standard theory" of gravitation, Einstein's General
Relativity (GR) has  passed many tests  within Solar system. To
reconcile the successful GR predictions within the solar system,
the extended gravity theories may be required to be scale sensitive. It
could be challenging and profound to locate the additional
curvature terms in our above discussions what form of scale dependence.

\textbf{Acknowledgements}

The insightful comments by three referees are highly appreciated that solidify this work considerably.
We would  especially like to thank Professor Sergei Odintsov too for his
careful reading this manuscript and many very helpful advice,
which have improved this paper a lot. Specially, he informed us the
non-unitarity problem in $R^2$ gravity and suggested reconsidering
this problem in Palatini formalism. We would also like to thank
Professors Sean Carroll, \'{E}anna Flanagan, Nadeem Haque,
Shin'ichi Nojiri,  Mark Trodden and  A. A Starobinsky for helpful
discussions and Professors Mauro Francaviglia and Igor Volovich
for helping us finding their earlier works. One of the
authors (XHM) would also like to express his thanks to the Physics Department of UoA for 
hospitality extended to him over where this paper is finally completed and Xinmin Zhang for lots of 
discussions. This work is partly
supported by China NSF, Doctoral Foundation of National Education
Ministry and ICSC-World Laboratory Scholarship.

\begin{appendix}
\end{appendix}

\end{document}